\documentclass{article}
\usepackage[dvips]{graphicx}

\title{Compression and Diffusion: A Joint Approach to Detect Complexity}

\begin{document}

\author{P. Allegrini$^{1,4}$, V. Benci$^{2,3}$, P.
Grigolini$^{4,5,6}$, P.Hamilton$^{7}$,\\ M. Ignaccolo$^{4}$,
G. Menconi$^{2,3}$, L. Palatella $^{5}$, G. Raffaelli$^{8}$, \\
N. Scafetta$^{4}$, M.  Virgilio$^{5}$, J.Yang$^{4}$} 
\maketitle
\small
\begin{center}
{\em $^{1}$ Istituto di Linguistica Computazionale del Consiglio
Nazionale delle}\\ 
{\em Ricerche, Area della Ricerca di Pisa, Via Alfieri 1,
San Cataldo, 56010, Ghezzano-Pisa, Italy}\\
{\em $^{2}$Dipartimento di Matematica Applicata, Universit\`{a} di
Pisa, Via Bonanno 26/b, 56127 Pisa, Italy}\\ 
{\em $^{3}$ Centro Interdisciplinare per lo Studio dei Sistemi Complessi, Universit\`{a}
di Pisa, Via Bonnanno, 25/b 56126 Pisa, Italy}\\
{\em $^{4}$Center for Nonlinear Science, University of North Texas, P.O. Box 311427, Denton, Texas 76203-1427 }\\
{\em $^{5}$Dipartimento di Fisica dell'Universit\`{a} di Pisa and
INFM Piazza Torricelli 2, 56127 Pisa, Italy }\\
{\em $^{6}$Istituto di Biofisica del Consiglio Nazionale delle
Ricerche,}\\ 
{\em Area della Ricerca di Pisa, Via Alfieri 1, San Cataldo,
56010, Ghezzano-Pisa, Italy }\\ 
{\em $^{7}$ Center for Nonlinear
Science, Texas Woman's University, P.O. Box 425498, Denton, Texas
76204}\\
{\em $^{8}$ International School
for Advanced Studies, Via Beirut 4, 34014 Trieste, Italy}
\end{center}
 \normalsize




\begin{abstract}
The adoption of the Kolmogorov-Sinai (KS) entropy is becoming a
popular research tool among physicists, especially when applied to a
dynamical system fitting the conditions of validity of the Pesin
theorem.  The study of time series that are a manifestation of
system dynamics whose rules are either unknown or too complex for a
mathematical treatment,  is still a challenge since the KS entropy is
not computable, in general, in that case. Here we present a plan of
action based on the joint action of two procedures,  both related to
the KS entropy, but compatible with computer implementation through
fast and efficient programs. The former procedure, called Compression
Algorithm Sensitive To Regularity (CASToRe), establishes the amount of
order by the numerical evaluation of algorithmic compressibility. The
latter, called Complex Analysis of Sequences via Scaling AND
Randomness Assessment (CASSANDRA), establishes the complexity degree 
through the numerical evaluation of the strength of an anomalous effect. This is the
departure, of the diffusion process generated by the observed
fluctuations, from ordinary Brownian motion. The CASSANDRA algorithm
shares with CASToRe a connection with the Kolmogorov complexity.
This makes both algorithms  especially suitable to study the transition
from dynamics to thermodynamics, and  the case of non-stationary
time series as well. The benefit of the joint action of these two
methods is proven by the analysis of artificial sequences with the
same main properties as the real time series to which the joint use of
these two methods will be applied in future research work.
\end{abstract}
\section{Introduction}
The KS entropy is a theoretical tool widely used by
physicists for an objective assessment of
randomness\cite{dorfman,beck}.  We consider a sequence of symbols
$\omega_{i}$, and we use a moving window of size $l$, with $l$ being
an integer number. This means that we can accommodate within this
window $l$ symbols of the sequence, in the order they appear, and that
moving the window along the sequence we can detect combinations of
symbols that we denote by $\omega_{0}
\omega_{1}\ldots\omega_{l-1}$. In principle, having available an
infinitely long sequence and a computer with enough memory and
computer time, for any combination of symbols we can evaluate the
corresponding probability $p(\omega_{0},
\omega_{1},\ldots\omega_{l-1})$.  Consequently, the $l$-th order empirical
entropy is
\begin{equation}
     \label{shannonentropy}
     H_l =
-\sum_{\omega_{0} \omega_{1}\ldots\omega_{l-1} }
p(\omega_{0}, \omega_{1},\ldots\omega_{l-1})
\ln (p(\omega_{0}, \omega_{1},\ldots\omega_{l-1})).
\end{equation}
The KS entropy is defined as
\begin{equation}
     \label{standarddefinition}
     h_{KS}= lim_{l \rightarrow \infty} \frac{H_l}{l}.
     \end{equation}
     This way of proceeding is of only theoretical interest since the
    KS  computer evaluation can hardly exceed a window length of
    order $10$ \cite{marcobuiatti}. The reason why the KS is so popular depends on the
    Pesin theorem \cite{dorfman,beck}. In the specific case where the symbolic
    sequence is generated by a well defined dynamic law, the Pesin
    theorem affords a practicable criterion to evaluate the KS
    entropy. To make an example, let us assume that the dynamic law is
    expressed by
    \begin{equation}
        \label{dynamiclaw}
        x_{n+1} = \Pi(x_{n}),
        \end{equation}
        in the interval $[0,1]$, where $\Pi$ has a derivative $\Pi '$.
        Using the Pesin theorem\cite{dorfman,beck} we can express
        $h_{KS}$ under the form
        \begin{equation}
	   h_{KS} = \int_{0}^{1}ln |\Pi'(x)| \rho(x) dx,
	   \label{pesin}
	   \end{equation}
	   where $\rho(x) dx$ denotes the invariant measure whose existence is
	   essential for the Pesin theorem to work.

In practice, the statistical analysis of a system of sociological,
biological and physiological interest, is done through the study of a
time series. This is a seemingly erratic sequence of numbers, whose
fluctuations are expected to mirror the complex dynamics of the system
under study. In general, we are very far away from the condition of
using any dynamic law, not to speak of the one-dimensional picture of
Eq. (\ref{dynamiclaw}). Thus the randomness assessment cannot be done
by means of Eq. (\ref{pesin}).  Furthermore, there is no guarantee
that a real time series refers to the stationary condition on which
the Pesin theorem rests. This might mean that the invariant
distribution does not exist, even if the rules behind the complex
system dynamics do not change with time. The lack of an invariant
distribution might be the consequence of an even worse condition: this
is when these rules do change with time. How to address the
statistical analysis of a real process, in these cases?

The main purpose of this paper is to illustrate the main ideas to
bypass these limitations. This is done with the joint use of two
techniques, both related in a way that will be discussed in this
paper, to the KS entropy. Both methods serve the purpose of making the
KS entropy computable, in either a direct or indirect way. The first
method is a compression algorithm.  This method applies directly to
the time series under study and evaluates its computable content. The
second method, on the contrary, does not refer directly to the data of
the time series, but it interprets them as diffusion generating
fluctuations, and then evaluates the entropy of the resulting
diffusion process, hence the name of Diffusion Entropy (DE) method. An
interesting goal of this paper is also to illustrate the connections
between these two methods.

The outline of the paper is as follows. In Section II we illustrate
the compression algorithm method. In Section III we illustrate the DE
method. Section IV is devoted to illustrating the dynamical model that
we use to create the artificial sequences. 
In Section V we show CASSANDRA  in action in
the case of sequences mimicking non-stationary processes. In Section VI we address the central core of this paper, the joint use of CASSANDRA and CASToRE. Section VII
illustrate the two methods in action on time series
characterized by a mixture of \emph{events} and \emph{pseudo events}.  What we mean by events and pseudo events will
be explained in detail in Section VII. Here we limit ourselves to saying that
by event we mean the occurence of a unpredictable fact, while the concept of pesudoevent implies predictability. In Section VIII,
using the logistic map at the chaos threshold, we show that the power
law sensitivity to initial condition leads to a form of localization
with the DE not exceeding a given upper limit. A balance on the results     
obtained is made in Section IX.
                
\section{The Computable Information Content method}
In the first of the two approaches illustrated in this paper the basic
notion is the notion of \textit{information}. Given a finite string
$s$ (namely a finite sequence of symbols taken in a given alphabet),
the intuitive meaning of \textit{quantity of information} $I(s)$
contained in $s$ is the length of the smallest binary message from
which we can reconstruct $s$.  This concept is expressed by the notion
of Algorithmic Information Content ($AIC$). We limit ourselves to
illustrating the main idea with arguments, intuitive but as close as
possible to the formal definition (for further details, see
\cite{Licata} and related references). We can consider a partial
recursive function as a computer $C$ which takes a program $p$ (namely
a binary string) as an input, performs some computations and gives a
string $s=C(p)$, written in the given alphabet, as an output. The
$AIC$ of a string $s$ is defined as the shortest binary program $p$
which gives $s$ as its output, namely
\[
I_{AIC}(s, C)=\min\{|p|:C(p)=s\},
\]
where $|p|$ means the length of the string $p$.
From this point of
view, the shortest program $p$ which outputs the string $s$ is a sort
of optimal encoding of $s$. The information that is necessary to
reconstruct the string is contained in the program. Unfortunately, this
coding procedure cannot be performed on a generic string by any
algorithm: the Algorithmic Information Content is not
computable by any algorithm (see Chaitin theorem in \cite{chaitin}).

                               Another measure of the information
content of a finite string can also be defined by a loss-less data
compression algorithm $Z$ satisfying some suitable properties which we
shall not specify here. Details are discussed in \cite{Licata}. We can
define the information content of the string $s$ as the length of the
compressed string $Z(s),$ namely,
\[
I_{Z}\left(  s\right)  =\left|  Z(s)\right|  .
\]
The advantage of using a compression algorithm lies in the fact that, this
way, the information content $I_{Z}\left(  s\right)$ turns out to be a
computable function. For this reason we shall call it Computable Information
Content $(CIC)$. In any case, given any string $s,$ we assume to have defined
the quantity $I(s)$ via $AIC$ or via $CIC.$
If $\omega$ is an infinite string, in general, its information is infinite;
however it is possible to define another notion: the complexity. The
complexity $K(\omega)$ of an infinite string $\omega$ is the average
information $I$ contained in a single digit of $\omega$, namely,
\begin{equation}
K(\omega)={\limsup}_{n\rightarrow\infty}\frac{I(\omega^{n})}{n}
\label{one},
\end{equation}
where $\omega^{n}$ is the string obtained taking the first $n$
elements of $\omega.$ If we equip the set of all infinite strings
$\Omega$ with a probability measure $\mu$, the couple $(\Omega,\mu)$
can be viewed as an information source, provided that $\mu$ is
invariant under the natural shift map $\sigma$, which acts on a string
$\omega=(\omega _i)_{i\in\mathbf{N}}$ as follows:
$\sigma(\omega)=\tilde{\omega}$ where $\tilde{\omega}_i=\omega_{i-1}\
\forall i\in\mathbf{N}$. The entropy $h_{\mu}$ of $(\Omega,\mu)$ can be
defined as the expectation value of the complexity:
\begin{equation}
h_{\mu}=\int_{\Omega}K(\omega)\;d\mu\ . \label{two}%
\end{equation}
If $I\left(  \omega\right)  =I_{AIC}\left(  \omega\right)  $ or $I\left(
\omega\right)  =I_{Z}\left(  \omega\right)  ,$ under suitable assumptions on
$Z$ and $\mu,$ $h_{\mu}$ turns out to be equal to the Shannon entropy. Notice
that, in this approach, the probabilistic aspect does not appear in the
definition of information or complexity, but only in the definition of entropy.

Chaos, unpredictability and instability of the behavior of dynamical
systems are strongly related to the notion of information. The KS
entropy illustrated in Section I can be interpreted as the average
measure of information that is necessary to describe a step of the
evolution of a dynamical system. As seen in Section I, the traditional
definition of KS entropy is given by the methods of
pro\-ba\-bi\-listic information theory: this corresponds to the a
version of the Shannon entropy adapted to the world of dynamical
systems.

We have seen that the information content of a string can be defined either
with probabilistic methods or using the $AIC$ or the $CIC$. Similarly, also the
KS entropy of a dynamical system can be defined in different ways. The
probabilistic method is the usual one, the $AIC$ method has been introduced by
Brudno \cite{brud}; the $CIC$ method has been introduced in \cite{gal4} and
\cite{ZIPPO}. So, in principle, it is possible to define the entropy of a
\emph{single} orbit of a dynamical system (which we shall call, as sometimes it
has already been done in the literature, \textit{complexity of the orbit}).
There are different ways to do this (see \cite{brud}, \cite{gal2},
\cite{Gaspard}, \cite{bonanno} , \cite{gal3}). In this paper, we consider a
method which can be implemented in numerical simulations. Now we shall describe
it briefly.

Using the usual procedure of symbolic dynamics, given a partition $\alpha$ of
the phase space of the dynamical system $\left(  X,\mu, T\right)  $, it is
possible to associate a string $\Phi_{\alpha}\left(  x\right)  $ to the orbit
having $x$ as initial condition. If $\alpha=(A_{1},\dots,A_{l}),$ then
$\Phi_{\alpha}\left(  x\right)  =(s_{0},s_{1},\dots,s_{k},\dots)$ if and only if%

\[
T^{k}x\in A_{s_{k}} \quad\forall\ k \ .
\]
If we perform an experiment, the orbit of a dynamical system can be described
only with a given degree of accuracy related to the partition of the phase
space $X$. A more accurate measurement implies a finer partition
of $X.$ The symbolic orbit $\Phi_{\alpha}\left(  x\right)  $ is a mathematical
idealization of these measurements. We can define the complexity $K(x,\alpha)$
of the orbit with initial condition $x$ with respect to the partition $\alpha$
in the following way
\[
K(x,\alpha)={\limsup}_{n\rightarrow\infty}\frac{I(x,\alpha,n)}{n}, %
\]
where
\begin{equation}
I(x,\alpha,n):=I(\Phi_{\alpha}\left(  x\right)  ^{n}).\label{info}
\end{equation}
Here $\Phi_{\alpha}\left(  x\right)  ^{n}$ represents the first $n$ digit of
the string $\Phi_{\alpha}\left(  x\right)  .$ Letting $\alpha$ vary among
all the \emph{computable partitions}, we set
\[
K(x)=\sup_{\alpha}K(x,\alpha)\ .
\]
The number $K(x)$ can be considered as the average amount of information
necessary to ''describe'' the orbit in the unit time when we use a
sufficiently accurate measurement device.

The notion of "computable partition" is based on the idea of
computable structure which relates the abstract notion of metric space
with computer simulations. The formal definitions of computable
partition are given in \cite{gal1}. We limit ourselves to describing
its motivation. Many models of the real world use the notion of real
numbers or more in general the notion of complete metric spaces. Even
if we consider a very simple complete metric space, as, for example,
the interval $\left[ 0,1\right] $, we note that it contains a
continuum of elements. This fact implies that most of these elements
(numbers) cannot be described by any finite alphabet.  Nevertheless,
in general, the mathematics of complete metric spaces is simpler than
the ''discrete mathematics'' in making models and the relative
theorems.  On the other hand, the discrete mathematics allows to make
computer simulations. A first connection between the two worlds is
given by the theory of approximation. But this connection becomes more
delicate when we need to simulate more sophisticated objects of
continuum mathematics. For example, an open cover or a measurable
partition of $\left[ 0,1\right]$ is very hard to simulate by
computer; nevertheless, these notions play a crucial role in the definition
of many quantities as, e. g., the KS entropy of a dynamical system or
the Brudno complexity of an orbit. For this reason, in the earlier
work leading to the foundation of the first method of this paper, the
notion of ''computable  structure" was  introduced. This is a new way to
relate the world of continuous models with the world of computer
simulations. The intuitive idea of computable partition is the
following one:
\begin{center}
a partition is computable if it can be recognised\\
by a computer.\end{center}
For example, an interval $\left[ a,b\right] $  belongs to a
computable partition if both $a$ and $b$ are computable real
numbers\footnote{A computable number is a real number whose binary
expansion can be given at any given accuracy by an algorithm.}. In
particular, a computable partition contains only a enumerable number
of elements.

In the above construction, the complexity of each orbit $K(x)$ is
defined independently of the choice of an invariant measure. In the
compact case, if $\mu$ is an invariant measure on $X$ then
$\int_{X}K(x)\;d\mu$ equals the KS entropy. In fact, in
\cite{Licata} it has been proved the following result.\\[10pt] {\bf
Theorem.} {\it If $(X,\mu,T)$ is a dynamical system on a compact space
and $\mu$ is ergodic, then for $\mu$-almost each $x\in X$ it holds:}
\begin{equation}
K_Z(x)=K_{AIC}(x)= h_{\mu}(T)\ \ .
\end{equation}
In other words, in an ergodic dynamical system, for almost all points
$x\in X,\;$ and for suitable choice of the partition
$\alpha,\;I(x,\alpha ,n)\sim h_{\mu}n.$ Notice that this result holds
for a large class of Information functions $I$ as for example the
$AIC$ and the $CIC.$ Thus we have obtained an alternative way to
understand the meaning of the KS entropy.

The above construction makes sense also for a \emph{non stationary system}.
Its average over the space $X$ is a generalization of the KS entropy to the
non stationary case. Moreover, the asymptotic behavior of $I(x,\alpha,n)$
gives an invariant of the dynamics which is finer than the KS entropy and is
particularly relevant when the KS entropy is null.

It is well known that the KS entropy is related to the
instability of the orbits. The exact relations between the KS entropy and the
instability of the system is given by the Pesin theorem. We shall recall
this theorem in the one-dimensional case. Suppose that the average rate of
separation of nearby starting orbits is exponential, namely,
\[
\Delta x(n)\simeq \Delta x(0)^{\lambda n}\;\;\;\mathrm{for }\Delta x(0)\ll 1,
\]
where $\Delta x(n)$ denotes the distance of these two points at time $n$. The number
$\lambda$ is called Lyapunov exponent; if $\lambda>0$ the system is unstable
and $\lambda$ can be considered a measure of its instability (or sensitivity
 to the initial conditions). The Pesin theorem implies
that, under some regularity assumptions, $\lambda$ equals the KS entropy.

There are chaotic dynamical systems whose entropy is null: usually they are
called weakly chaotic. Weakly chaotic dynamics appear in the field of 
self-organizing systems, anomalous diffusion, long- range interactions and many
others. In such dynamical systems the amount of information necessary to
describe $n$ steps of an orbit is less than linear in $n$, then the KS
entropy is not sensitive enough to distinguish the various kinds of weakly
chaotic dynamics. Nevertheless, using the ideas we illustrate in this section, the
relation between initial data sensitivity and information content of the
orbits can be extended to these cases.

To give an example of such a generalization, let us consider a
dynamical system $([0,1],T)$ where the transition map $T$ is
\textit{constructive\footnote{A constructive map is a map that can be
defined using a finite amount of information, see \cite{gal3}.}}, and
the function $I(x,\alpha,n)$ is defined using the $AIC$ in a slightly
different way than before (use open coverings instead of partitions,
see \cite{gal3}). If the speed of separation of nearby starting orbits
goes like $\Delta x(n)\simeq\Delta x(0)f(x,n)$, then for almost all
the points $x\in\lbrack0,1]$ we have
\begin{equation}
\label{chaos}
I(x,\alpha,n)\sim\log(f(x,n)).
\end{equation}
In particular, if we have power law sensitivity ($\Delta x(n)\simeq\Delta x(0)n^{p}$), the information content of the orbit is
\begin{equation}I(x,\alpha,n)\sim
p\log(n)\ .\label{utile}\end{equation} If we have a stretched
exponential sensitivity ( $\Delta x(n)\simeq\Delta x(0)2^{\lambda
n^{p}}$, $p<1$) the information content of the orbits will increase
with the power law: 
\begin{equation}
I(x,\alpha,n)\sim n^{p} .
\label{utile2}
\end{equation}
Since we have shown that the analysis of $I(x,\alpha,n)$
gives useful information on the underlying dynamics and since
$I(x,\alpha,n)$ can be defined through the $CIC$ methodology, it turns out that it
can be used to analyze experimental data using a compression algorithm
which is efficient enough  and which is fast
enough to analyze long strings of data. We have implemented a
particular compression algorithm we called CASToRe: Compression
Algorithm Sensitive To Regularity (\cite{ZIPPO}).

CASToRe is an encoding algorithm based on an adaptive dictionary and
it is a modification of the LZ78 algorithm. Roughly speaking, this
means that it translates an input stream of symbols (the file we want
to compress) into an output stream of numbers, and that it is possible
to reconstruct the input stream knowing the correspondence between
output and input symbols. This unique correspondence between sequences
of symbols (words) and numbers is called \emph{the dictionary}. The
dictionary is adaptive because it depends on the file under
compression, in this case the dictionary is created while the symbols
are translated.  

At the beginning of encoding procedure, the dictionary is empty. In
order to explain the principle of encoding, let us consider a point
within the encoding process, when the dictionary already contains some
words.  The algorithm starts analyzing the stream, looking for the
longest word W in the dictionary matching the stream. Then it looks
for the longest word Y in the dictionary where W + Y matches the
stream. The new word to add to the dictionary would be W +
Y. More details on its internal running are described in the Appendix
of \cite{Licata}.  In the following, the numerical results addressed
to the CIC method have been performed using the algorithm CASToRe.

\section{The DE method}
Let us now illustrate the second of the two methods of analysis of this paper. This method
rests on DE used for the first time in Ref. \cite{nic1}. This method turned out to be a very efficient and reliable 
way to determine scaling \cite{giacomo}. Here, after a short review of the DE method, we argue that the DE is not only a method
of scaling detection and that with the use of two windows it can be expressed in a form
that turns out to be very efficient to study the case of rules changing with time.

The first step
of the DE method is the same as that of the pioneering work of
Refs.\cite{detrendingindna,detrendinginheartbeating}. This means that
the experimental sequence is converted into a kind of Brownian-like
trajectory. The second step aims at deriving many distinct diffusion
trajectories with the technique of moving windows of size $l$.
The
reader should not confuse the mobile window of size $l$ with the mobile window
of size $L$ that
will be used later on in this paper to detect non-stationary properties.
For this reason we shall refer to the mobile windows of size $L$ as
\emph{large windows}, even if the size of $L$ is relatively small, whereas
the mobile windows of size $l$ will be called \emph{small windows}.
The large mobile window  has to be interpreted as a sequence on its own,
with local statistical properties to reveal, and will be analyzed by means
of small windows of size $l$, with $l < L$. The success of the method depends
on the fact that the DE makes a wise use of the statistical
information available. In fact, the small windows overlap and are
obtained by locating their left border on the first site of the
sequence, on the second second site, and so on. The adoption of
overlapping mobile windows is dictated by the wish  of establishing a
connection with the KS  \cite{beck,dorfman} method, and it has
the effect of creating many more trajectories than the Detrended
Fluctuation Analysis (DFA)\cite{detrendingindna,detrendinginheartbeating}.

In conclusion, we create a conveniently large number of trajectories by
gradually moving the small window from the first position, with the left border
of the small window coinciding with the first size of the sequence,
to the last position, with the right border of the small window
coinciding with the last site of the sequence:
$$
x_k(l)=\Sigma_{s=k}^{k+l-1}\xi(s)\ , \ \ \  0<k<N-l+1\ .
$$
After this stage, we
utilize the resulting trajectories, all of them with the initial position
located at $x = 0$, to produce a probability distribution at
``time'' $l$:
$$
p(x,l)\ =\ \Sigma_{k=0}^{N-l+1}\delta_{x_k(l),x} ,
$$ 
where $\delta_{i,j}$ denotes the delta of Kronecker.

According to a  tenet of the Science of Complexity \cite{yaneer,mandelbrot}, complexity is related to the concept of diffusion scaling.
Diffusion scaling is defined by
  \begin{equation}
	p(x,l) = \frac{1}{l^{\delta}} F(\frac{x}{l^{\delta}}).
	\label{scalingdefinition}
	\end{equation}
Complex systems are expected to generate a departure from the condition of ordinary diffusion,
where $\delta = 0.5$ and $F(y)$ is a Gaussian function of $y$. For this reason one of the goals of the statistical analysis of time series \cite{detrendingindna,detrendinginheartbeating}
is the determination of $\delta$. The methods currently adopted measure the second moment of the diffusion process, $< x^{2}>  \propto t^{2H}$.
However, the validity of this way to detect scaling is questionable since the parameter $H$, usually called Hurst coefficient \cite{detrendingindna,detrendinginheartbeating}, is known to coincide with
$\delta$ only in the case of fractional Brownian motion \cite{mandelbrot}. It is known that not always the scaling of the second moment
is a fair representation of the property of Eq. (\ref{scalingdefinition}). For instance, the work on dynamic approach to L\'{e}vy 
processes \cite{allegro}
shows that that the second moment yields a scaling different from the scaling corresponding to the definition of Eq. (\ref{scalingdefinition}). 
The DE method, on the contrary, if the condition of Eq. (\ref{scalingdefinition}) is fulfilled, yields the correct value for the scaling parameter $\delta$.
Let us see why. The Shannon entropy of the diffusion process reads
\begin{equation}
S(l) = - \int_{\infty}^{+\infty} p(x,l) \ln p(x,l) dx.
\label{shannon}
\end{equation}
Let us plug Eq. (\ref{scalingdefinition}) into Eq. (\ref{shannon}). After some trivial change of integration variable we get
\begin{equation}
     S(l) = A + \delta \ln (l),
     \label{shapeofdiffusionentropy}
     \end{equation}
where $A$ is a constant, whose explicit form is of no interest for the current discussion. From Eq. (\ref{shapeofdiffusionentropy}) we derive a natural way to
measure the scaling $\delta$ without  recourse to
     any form of detrending. 

However, it is worth pointing out that the $DE$ is not only a way to detect scaling. It is much more than that, and one of the main purposes of this paper
is to illustrate these additional properties with the help of the perspective of the computable information content, illustrated in Section II. To be more specific,
we shall study the DE in action in the case of diffusion processes generated by a dynamic process with either the stretched exponential sensitivity of Eq. (\ref{utile2}) or the power law sensitivity
of Eq. (\ref{utile}). We shall see that in those specific cases the DE method of analysis allows us to shed light into the fascinating process of transition from dynamics
to thermodynamics that is made to become visible when those anomalous conditions apply. We shall see also that the time size of the process of transition from dynamics to thermodynamics
is enhanced by the superposition of a random  and deterministic process. We shal illustrate this property with an apparently trivial dynamical model. This is the case where the time series under study consists of the sum of a regular motion (a harmonic oscillation) and a completely random process.
We shall discuss also the case where the harmonic oscillation is replaced by a quasi-periodic process yielding the power law sensitivity of Eq. (\ref{utile}). 

In this paper we shall be dealing with two different forms of non-stationarity. The former has to do with fixed dynamic rules, which are, however, incompatible with the
existence of an invariant distribution. An example of this kind is given by the Manneville map with $z > 2$, and with the dynamic model of Section IV. We recall that a symbolic sequence
$\{\sigma_i\}$ is stationary if the frequency of each symbol tends to
a limit which is different from zero. If we associate a symbolic
sequence to an autonomous dynamical system, that sequence may also tend to a vanishing value, thereby implying non-stationarity, in the first sense. 
The second type of non-stationarity refers to the case when the dynamic rules change upon time change.  This has to do with a case that is
frequently met, when studying time series (see, for instance,
Ref.\cite{nic1}).  In this specific case, we are led
to select a portion of the time series, located, for instance, at two distinct specific times. The purpose of the statistical analysis should be 
to assess if the statistical properties, and consequently the rules responsible
for these statistical properties,  at these two distinct times are different or not. Since
the statistical properties might be time dependent,  the portion of sequence to
examine cannot be too large. On the other hand, if the size of that
portion is relatively small, there is no time for the process to reach
the scaling regime. It is precisely with this picture in mind that the
authors of Ref. \cite{patti2} designed a special version of the DE
method that they called Complex Analysis of Sequences via Scaling AND
Randomness Assessment (CASSANDRA)\cite{patti2}. Actually, the
connection with the non-stationary aspects of the time series under
study does not emerge from the literal meaning of the acronym, but
rather from the suggestion of the Cassandra myth. This means indeed
that the algorithm with the same name as the daughter of Priam and
Ecuba is expected to be useful to make prediction on the basis of the
assessment of how rules change upon function of time. The idea is
that catastrophic events might be preceded by a preliminary slow
process with the rules slightly changing in time. It is this slight
change of rules that CASSANDRA should detect for prediction
purposes.

\section{Dynamical Model for Intermittence}
Let us consider a simple dynamical model generating events. By event
we mean something completely unpredictable. The example here under
discussion will clarify what we really mean.  Let us consider the
equation of motion
\begin{equation}
\label{sporadicrandomnessmodel}
\dot{x}=\lambda\ x^z\ ,\ z>1,
\end{equation}
where the variable $x$ moves within the interval $I \equiv [0,1]$.
The particle with this coordinate moves from the left to the
right. When this particle reaches the right border of the interval $I$
is injected back to a new initial condition within the interval $I$,
selected by means of a random number generator.  This is an event. Let
us imagine that the exit times are recorded. Let us assume that the
series under examination is $\{t_{i}\}$, the set of these exit
times. Since these exit times are determined by randomly chosen
initial conditions, which are not recorded, they can be considered as
being manifestations of events. In Section 6  we
shall study the case when the set $\{t_{i}\}$ is a mixture of events
and pseudo events, and this will serve the purpose of illustrating in
a deeper way our philosophical perspective. For the time being, let us
limit ourselves to studying a sequence of events. Let us change the
sequence $\{t_{i}\}$ into the new sequence $\{\tau_{i}\}$, where
\begin{equation}
     \label{new sequence}
     \tau_{i} \equiv t_{i+1} - t_{i}.
\end{equation}
With very simple arguments similar to those used in \cite{aquino}, and
in the earlier work of Ref. \cite{geisel} as well, we can prove that
the probability distribution of these times is given by
\begin{equation}
\label{waitingtimes}
     \psi(\tau) = (\mu -1) \frac{T^{\mu -1}}{(T + \tau)^{\mu}},
     \end{equation} with $\mu = z/(z-1)$ and $T = \frac{(\mu-1)}
     {\lambda}$.

In the special case where $2 < \mu $ the mean waiting time is given by
\begin{equation}
\label{mean}
<\tau> = \frac{T}{(\mu - 2)}.
\end{equation}
Using the arguments of Ref. \cite{annunziato} it can be shown that this dynamic model for $\mu > 3$ generates ordinary Gaussian diffusion, while 
for $\mu < 3$ (with $\mu >2$ still valid) yields L\'{e}vy statistics. From the entropic point of view, however, we do not observe any abrupt change when $\mu$ crosses the border between the L\'{e}vy and the Gauss basin of attraction. This corresponds to the general prescriptions of Section II, and it is also supported by intuitive arguments as follows.  We assume that the drawing of a random number implies an
entropy increase equal to $H_{E} $. It is evident that the rate of entropy
increase per unit of time is $H_{E}/{\tau}$, as it can be easily realized
by noticing that at a given time $t$ (very large), the number of
random drawings is equal to $t/<\tau>$. We note that this heuristic
form of entropy must not be confused with the KS entropy. For this
reason we use the subscript $E$ standing for
\emph{external}. This means that randomness has an external origin,
being dependent on the random drawing of initial condition.  The
available random number generators are not really random. However,
here for simplicity we adopt an idealization where the choice of
initial condition is really a random process, this being external to
the dynamics illustrated by Eq. (\ref{sporadicrandomnessmodel}). We see that this
form of entropy, as the KS entropy illustrated in Section II, does not result in any abrupt change when $\mu$ crosses the border 
between Gauss and L\'{e}vy statistics.  

The compression algorithm illustrated in Section II, as we have seen,
serves the purpose of making the KS entropy computable. Therefore, it
is convenient to derive an analytical expression for the KS
entropy. This can be done with heuristic arguments inspired to the
method proposed by Dorfmann\cite{dorfman}. We create a Manneville-like
map with a laminar region and a chaotic region. The key ingredient of
this map is given by the parameter $\lambda$, which is assumed to
fulfill the condition
\begin{equation}
\label{laminarlaminar}
\lambda << 1.
\end{equation}
This means that the integration step, equal to $1$, can be regarded as
infinitesimally small. This has important consequences. The original
Manneville map\cite{manneville} reads
\begin{equation}
\label{originalmanneville}
x_{n+1} = x_{n} + x_{n}^{z}, (mod. 1), z \leq 1.
\end{equation}
The region belonging to the interval $ I = [0,d]$, with $d$ defined by
$d + d^{z} = 1$, is called laminar region and is usually studied by
adopting a continuum time approximation. The Manneville map at $z = 1$
becomes identical to the Bernoulli shift map, and so to a coin tossing
process \cite{beck}. The use Eq.(\ref{sporadicrandomnessmodel}) with
the condition (\ref{laminarlaminar}) makes it possible for us to reach
the limiting condition $z = 1$ without ever leaving the continuous
time approximation on which Eq.(\ref{sporadicrandomnessmodel}) rests.
Using this model, the interval $I = [0,1]$ is divided two into
portions, the first ranging from $x = 0$ to $x = \epsilon$ (laminar
region) and the second from $ x = \epsilon$ to $x = 1$ (chaotic
region). The parameter $\epsilon$ is defined by the following condition
\begin{equation}
\label{definition}
(1-\epsilon) + \lambda (1-\epsilon)^{z} = 1 .
\end{equation}
The adoption of  condition (\ref{laminarlaminar}) makes the value
of $\epsilon$ become very close to that of $\lambda$. The chaotic
region is defined by the Bernoulli-like form
\begin{equation}
\label{chaoticregion}
x_{n+1} = \frac{x_{n} + \epsilon -1}{\epsilon}.
\end{equation} 
In the limiting case of extremely small $\lambda$'s the size of
chaotic region tends to zero, but its contribution to the Lyapunov
coefficient cannot be neglected, since it is much larger than the
Lyapunov coefficients of the laminar region. The invariant
distribution $p(x)$ is shown to be $p(x)= 1/x^{1/(\mu-1)}$. Thus, a
good approximation to the prescription of the Pesin
theorem \cite{dorfman,beck} is given by
\begin{equation}
\label{analyticalks}
h_{KS} = \frac{\mu -2}{\mu-1} \left[\int_{0}^{1}\frac{1}{x^{\frac{1}{(\mu -1)}}} \ln(1 + \frac{\mu}{T} x^{\frac{1}{\mu -1}})dx + \lambda \ln(\frac{1}{\lambda}) \right].
\end{equation}
In the case where the parameter $T$ is very large, we can take into
account only the first term of the Taylor expansion of the
logarithm. The new expression, written as a function of $z$, reads

\begin{equation}
\label{simpleanalyticalformula}
h_{KS} = (2-z) \lambda [ z + \ln (1/\lambda)].
\end{equation}

We see that when the mean waiting time becomes equal to infinity (at
$z = 2$, or $\mu =2$) the KS entropy vanishes,  a property shared by
both KS entropy and external entropy. When we explore $z > 2$ (or $\mu
<2$) we enter a regime, characterized by very interesting properties.
The adoption of the same heuristic arguments as those yielding the KS
expression of Eq. (\ref{analyticalks}) makes it possible to
 prove \cite{massi} that the external entropy increase as
$t^{\mu-2}$. The information content approach of Section II yields
equivalent results.  When $1<z<2$, the information content of a
trajectory with starting point $x$ and for $n$ steps is
\begin{equation}
I_{AIC}(x,\alpha,n)\sim K\ n\ , 
\end{equation} 
where the factor $K$ can be considered a numerical estimation of the
KS  entropy of the system $h_{KS}(T)$.   The table shows some numerical results for this case.  
\\[5pt] 
\begin{center}
\begin{tabular}{|c|c|c|c|}
\hline
{z value} & {$\lambda$ value} & {KS entropy }& {Complexity K }\\
\hline
\hline
1.8 &  0.1 & 0.082 & 0.079\\
1.5 & 0.04 & 0.0943 & 0.0936\\
1.2 & 0.025 & 0.0977 & 0.980\\
1.1 & 0.022 & 0.0980 & 0.993\\
\hline
\end{tabular}\label{tableKolm}
\end{center}
\vskip 0.5 truecm

When
$z>2$ the Manneville map is weakly chaotic and non stationary. It can
be proved \cite{Gaspard}, \cite{bonanno}, \cite{gal3} that for almost
each $x$ (with respect to the Lebesgue measure)
\begin{equation}
I_{AIC}(x,\alpha,n)\sim n^{\frac{1}{z-1}}\ .\label{ixan}
\end{equation}
Some numerical results about the exponent are the following:\\[5pt]
\begin{center}
\begin{tabular}
[c]{|c|c|c|}\hline
z & CIC & theoretical value\\\hline\hline
2.8 & 0.573 & 0.555\\\hline
3 & 0.533 & 0.5\\\hline
3.5 & 0.468 & 0.4\\\hline
\end{tabular}
\end{center}
\section{The Balance between Randomness and Order Changing with Time}
We address the problem of detecting non-stationary effects in time
series (in particular fractal time series) by means of the DE  method. This means that the experimental sequence under
study, of size $N$, is explored with a window of size $L << N$. The
artificial sequence under study is described by

\begin{equation}
\xi_{b}(t) = \kappa\xi(t) + A cos (\omega t).
\label{withbias}
\end{equation}
The second term on the right hand side of this equation is a
deterministic contribution that might mimic, for instance, the season
periodicity of Ref. \cite{nic1}.
The first term 
on the right 
hand side is a fluctuation with no correlation that can be correlated, 
or not,  to the harmonic bias.

Fig. 1 refers to the case when the random fluctuation has no
correlation with the harmonic bias. It is convenient to illustrate
what happens when $\kappa = 0$. This is the case where the signal is
totally deterministic, being reduced to $\xi_{b}(t) = A cos (\omega t)$. It would be nice if the entropy in this case
did not increase upon increasing $l$.  However, we must notice that
the method of mobile windows implies that many trajectories are
selected, the difference among them being a difference on initial
conditions. Entropy initially increases. This is due to the fact that
the statistical average on the initial conditions is perceived as a
source of uncertainty. However, after a time of the order of the
period of the deterministic process, a regression to the condition of
vanishing entropy occurs, and it  repeats at multiples of the period of the deterministic process. 
Another very remarkable fact is that the maximum
entropy value is constant, thereby signaling correctly that we are in
the presence of a periodic process: the initial entropy increase,
caused by the uncertainty on initial conditions, cannot keep going forever, and it is balanced by the
recurrences.

Let us now consider the effect of a non vanishing
$\kappa$. We see that the presence of an even very weak random
component makes an abrupt transition to occur from the condition where
the DE is bounded from above, to a new condition where
the recurrences are limited from below by an entropy increase
proportional to $0.5 \ln l$.  In the asymptotic time regime the DE method
yields, as required, the proper scaling $\delta =0.5$. However, we
notice that it might be of some interest for a method of statistical
analysis to give information on the extended regime of transition to
the final \emph{thermodynamic} condition. We notice that if the DE method is
interpreted as a method of scaling detection, it might also give the
impression that a scaling faster than the ballistic $\delta$ is
possible. This would be misleading. However, this aspect of the DE method,
if appropriately used, can become an efficient method to monitor the
non-stationary nature of the sequence under study, as we shall see with other examples in this section.

\begin{figure}
\begin{center}
\includegraphics[width=7cm,angle=270]{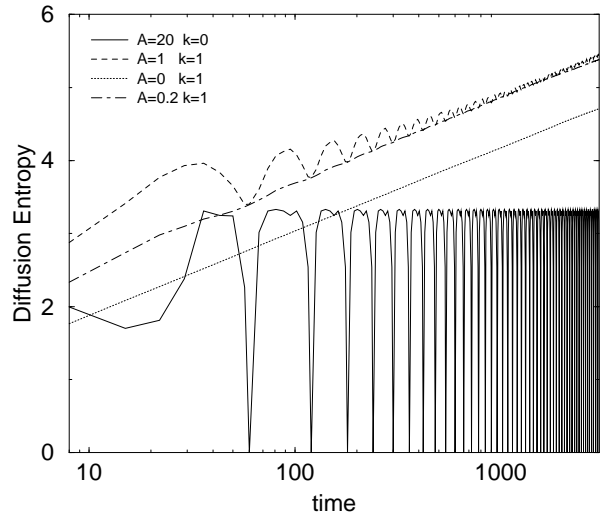}
\caption{The DE $S_{d}(l)$ as a 
function of time $l$ for 
different sequences of the type of Eq. (12).}
\end{center}
\label{fig1patti2} 
\end{figure}

\begin{figure}
\begin{center}
\includegraphics[scale=.40,angle=270]{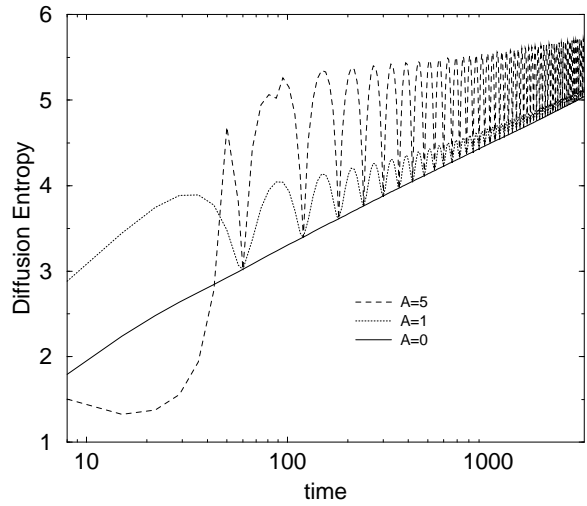}
\caption{\label{fig2patti2}
The DE  $S_{d}(l)$ as a function of time $l$ for 
different sequences of the type of Eq. (27) with the prescription of 
Eq. (28) for the random component.
} 
\end{center}
\end{figure}

In the special case where the fluctuation $\xi(t)$ is correlated 
to the bias, the numerical results illustrated in
Fig. \ref{fig2patti2} show that the time evolution of the DE is
qualitatively similar to that of Fig. 1.  The correlation
between the first and the second term on the right hand side of
Eq. (\ref{withbias}) is established by assuming

\begin{equation}
\xi(t) = \xi_{0}(t) cos (\omega t),
\label{modulatedintensity}
\end{equation}
where $\xi_{0}(t)$ is the genuine independent fluctuation, without
memory, whose intensity is modulated to establish a correlation with
the second term.  It is of some interest to mention what happens when
$A = 0, \kappa = 1$, and consequently $\xi_{b}(t)$ coincides with
$\xi(t)$ of Eq. (\ref{modulatedintensity}). In this case we get the
straight (solid) line of Fig. \ref{fig2patti2}.  This means that the
adoption of the assumption that the process is stationary yields a
result that is independent of the modulation.

\begin{figure}
\label{sinus}
\begin{center}
\includegraphics[scale=.40,angle=270]{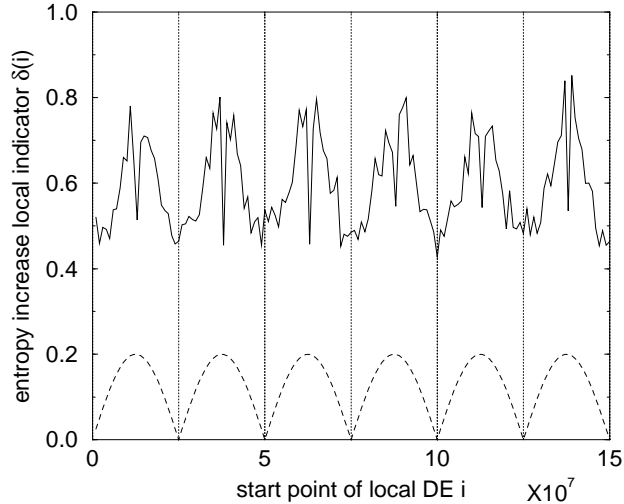}
\caption{\label{fig3patti2} The method of the two mobile windows applied to
a sequence given by Eq. (27.) with $A = 0$ and $\xi(t)$ given by
Eq. (28).  The dashed line represents the amplitude of the harmonic component (not in
scale) as a function of the position $i$ of the left border of the
large moving window. We selected $L = 512$ }
\end{center}
\end{figure}

We use this interesting case to illustrate the extension of the DE method. 
We note that the name CASSANDRA refers to this extension of the
DE method \cite{patti2}. As earlier mentioned, this
extension is based on the use of two mobile windows, one of length $L$
and the traditional one of length $l \ll L$.  This means that a large
window of size $L$, with $L \ll T=2\pi/\omega$, is located in a given
position $i$ of the sequence under study, with $i \leq N-L$, and the
portion of the sequence contained within the window is thought of as
being the sequence under study. We record the resulting $\delta$
(obtained with a linear regression method) and then we plot it as a
function of the position $i$. In Fig. \ref{fig3patti2} we show that this way
of proceeding has the nice effect of making the periodic nature of the process show up, 
in a condition where the adoption of small windows running over the whole sequence
would produce  the impression of the existence of a scaling regime.

\begin{figure}
\begin{center}
\includegraphics[width=7cm,angle=270]{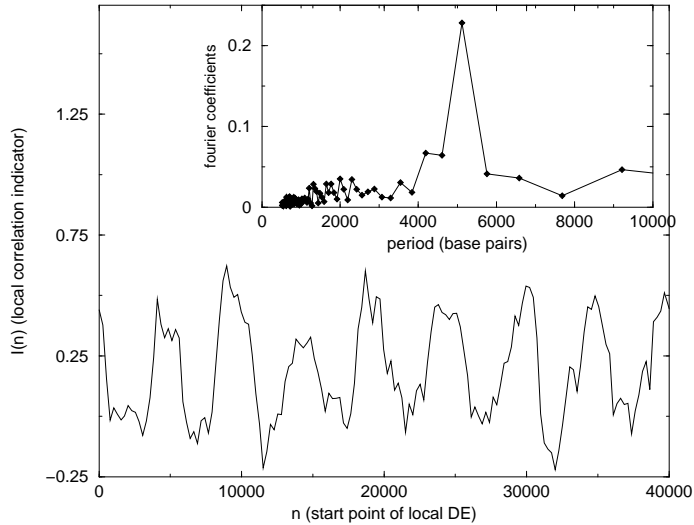}
\caption{\label{fig4patti2} The method of the two moving windows with
$l_{max}=30$ applied to the analysis of an artificial CMM sequence
with periodic parameter $\epsilon$. The period of the variation of
$\epsilon$ is $5000$ bps and the analysis is carried out with moving
windows of size $2000$ bps.  Inset: Fourier spectral analysis of
$I(n)$.}
\end{center}
\end{figure}

Let us now improve the method to face non-stationary condition even
further. As we have seen, the presence of time dependent condition
tends to postpone or to cancel the attainment of a scaling
condition. Therefore, let us renounce using Eq. (\ref{shannon}) and let us
proceed as follows. For any large mobile window of size $L$ let us
call $l_{max}$ the maximum size of the small windows. Let us call $n$
the position of the left border of the large window, and and let us
evaluate the following property

\begin{equation}
\label{cassandra}
I(n) \equiv \sum_{l=2}^{l_{max}} \frac{S_{d}(l) - [S_{d}(1)+0.5 \ln l]}{l}.
\end{equation}
The quantity $I(n)$ detects the deviation from the slope  that the DE would have in the random case.
Since in the regime of transition the entropy increase can be much
slower than in the corresponding random case, the quantity $I(n)$ can
also bear negative values. This indicator affords a satisfactory way
to detect local properties.  As an example, Fig. \ref{fig4patti2}  shows a case based on
the DNA model of Ref. \cite{maria} called Copying Mistake Map
(CMM). This is a sequence of symbols $0$ and $1$ obtained from the
joint action of two independent sequences, one equivalent to tossing a
coin and the other equivalent to establishing randomly a sequence of
patches whose length is distributed as an inverse power law with index
$\mu$ fitting the condition $2 < \mu < 3$.  The probability of using
the former sequence is $1 - \epsilon$ and that of using the latter is
$\epsilon$.  We choose a time dependent value of $\epsilon$:

\begin{equation}\label{modulatednoiseintensity}
\epsilon=\epsilon_0 [1-cos(\omega t)].
\end{equation}
The adoption of the DE method in the original version \cite{nic1}, namely with only small windows running over the whole sequence of data, would not reveal this periodicity. In Fig. \ref{fig4patti2} we show that, on the contrary, CASSANDRA, namely 
the DE method with two kinds of moving windows, makes it possible for us to distinctly perceive a fluctuation around the Brownian scaling, from regions with a scaling faster to regions with a scaling slower than the ordinary scaling $\delta = 1/2$. Of course, these scaling fluctuations do not refer to a genuine scaling changing with time, but rather to a processes of transition from dynamics to thermodynamics
with a changing rate. From Fig. 4 we see that, although partially random, this changing rate satisfactorily corresponds to the periodicity
of Eq. (30). 

\begin{figure}
\centerline{\includegraphics[angle=270,scale=.50]{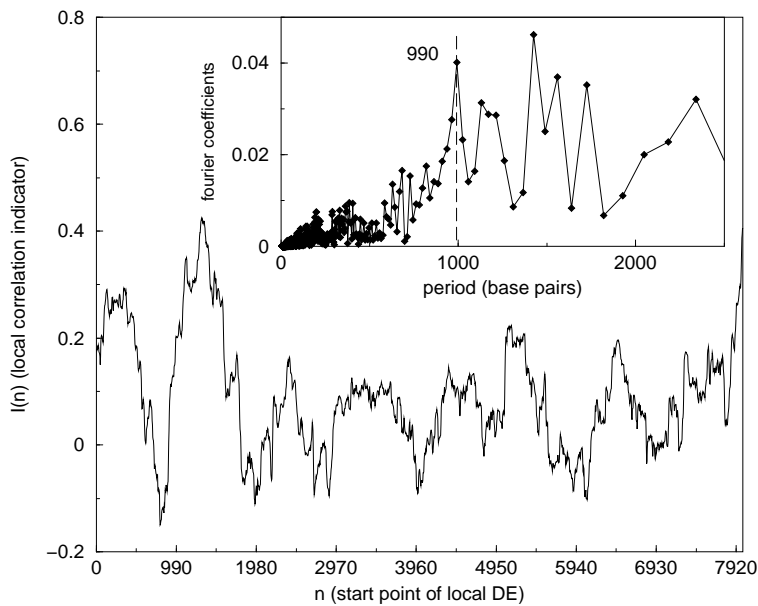}}
\caption{\label{fig5patti2} 
The method of two mobile windows applied to the analysis of the human 
DNA sequence. The method of two mobile windows ($l_{max}=20$
$L=512$) detects a periodicity 
of 990 bps. Inset: Fourier spectral analysis of $I(n)$. 
} 
\end{figure}

This paper, as earlier stated, is devoted to illustrating the benefits stemming
from the joint use of CASSANDRA and CASToRE by means of artificial sequences. 
However, before addressing the important issue of the connection between CASToRE and CASSANDRA, whose joint action on real sequences will be illustrated in future work,
we want to show CASSANDRA at work on real sequences.  Therefore, let us
address the problem of the search of hidden periodicities in DNA
sequences.  Fig. \ref{fig5patti2} shows a distinct periodic behavior
for the human T-cell receptor alpha/delta locus. A period of about 990
base pairs is very clear in the first part of the sequence (promoter
region), while several periodicities of the order of $1000$ base pairs
are distributed along the whole sequence.
These periodicities, probably due to DNA-protein interactions in
active eukaryotic genes, are expected by biologists, but the current
technology is not yet adequate to deal with this issue, either from
the experimental or from the computational point of view: such a
behavior cannot be analyzed by means of crystallographic or structural
NMR methods, nor would the current (or of the near future) computing
facilities allow molecular dynamics studies of systems of the order of
$10^6$ atoms or more. In conclusion, CASSANDRA (and to the close connection between the two methods, CASToRE as well) can be really adopted with success to analyze real sequences.

\section{The joint use of CASSANDRA and CASToRE }
We now address the issue that is the central core of this paper: the joint use of CASSANDRA and CASToRE. 
We compare the former to the latter method in a case similar to that of
Eq. (\ref{withbias}),  namely, the case  when we
have a periodic signal plus noise.  We have seen CASSANDRA in action
with real number signals. In order to use CASToRe it is necessary to
define a symbolic sequence, for instances,  $1$'s and $0$'s.
As a consequence of the theoretical arguments of Section II, concerning periodic or quasiperiodic signals, CASToRe is expected to yield a logarithmic increase of the CIC. We have seen in Fig. 1 that CASSANDRA in this case yields for  $S(t)$  saturation with periodical recurrences. In the case
where only randomness is present, the dynamic model of Section V with $\mu > 2$, we have seen that $S(t)$ yields the
linear increase with respect to the logarithm of time, $S(t) = S(0)+\delta \log(t)$. In this case, we know from  Section II  that  the CIC increases linearly in time.
We aim now at exploring these predictions with a close comparison between CASToRE and CASSANDRA.

\begin{figure}
\label{paologiulia1}
\begin{center}
 \includegraphics[width=16cm]{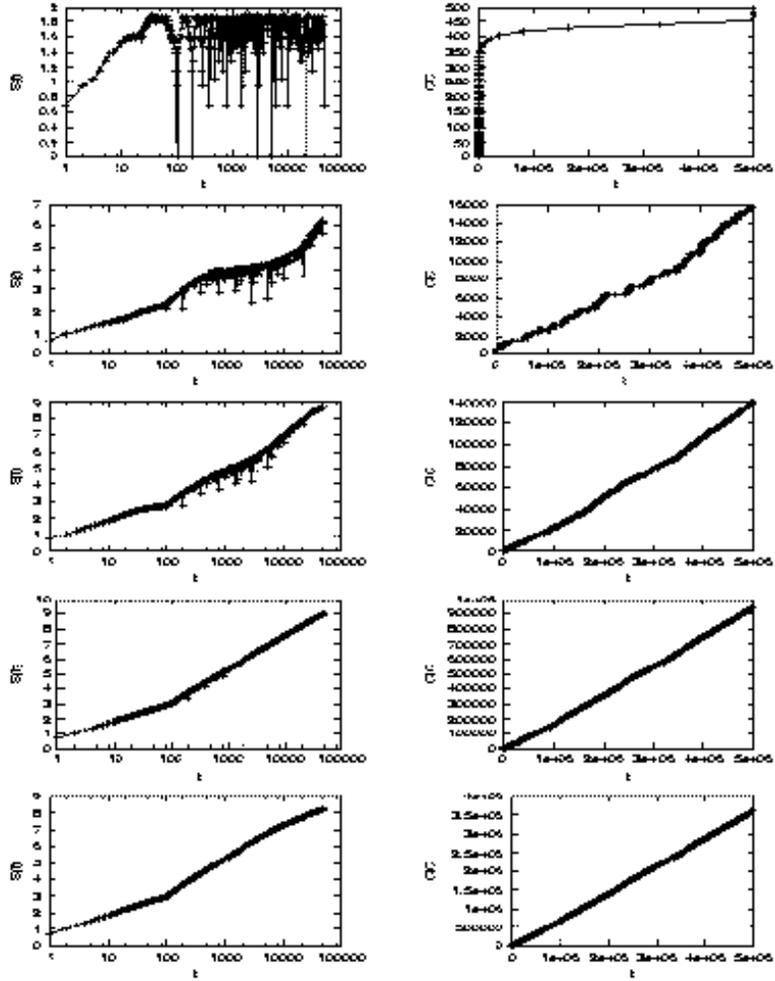}
\caption{Periodic signal plus noise (see text for details). The
figures in the first column refer to the CASSANDRA experiment, while the
ones in the second column refer to CASToRe. Figures on the same raw refer 
to the same value of $\epsilon$, and in particular, from top to bottom
$\epsilon=10^{-7},10^{-5},10^{-4},10^{-3},10^{-2}$.}
\end{center}
\end{figure}

Let us explain how to generate the first of the two sequences $\{\xi_{n}\}$ that will be used to show both CASToRE and CASSANDRA in action. First we  generate the sequence $\{\zeta_n\}$ as a periodic sequence repeating a random
pattern of 1's and 0's of length 100 for 50 times. Then, we build up the first sequence  $\{\xi_{n}\}$
following again the prescription of  the CMM of ref \cite{maria}, earlier used in this section. The periodic modulation of Eq. (30) will be used to build up the second sequence. Thus, we define the first sequence as:

\begin{equation}
\xi_n=\left\{ \begin{array}{cccc} 
\mbox{random}(0,1)& \mbox{with prob.} &\epsilon& \\
\zeta_n & \mbox{with prob.}&1-\epsilon,& \\
\end{array}
 \right.
\end{equation}
where  the first line stands for a random extraction of either the symbol $1$ or the symbol
$0$, by means of a fair coin tossing.
In Figs. 6 we illustrate the result of the comparison between the results of CASSANDRA and those of CASToRE, with the left and right columns being devoted to
CASSANDRA and CASToRE, respectively.  In the first line we illustrate the case when
no randomness is present. $S_d(t)$ increases for $ t< 100$, since there is no correlation for time distances smaller than the period $100$. Then it regresses back to $0$, after which it
increases and decreases again with periodic recurrences. This behavior is very close to that of Fig. 1.  
On the other hand, as far as CASToRE is concerned,  as explained in Section II, it takes some time 
for CIC to recognize the sequence periodicity. This means  that CASToRE has to store a given number of 
periods before yielding the logarithmic increase in time dicated by the periodic character of the sequence under study.
This expectation is fully confirmed  by the top figure of the right column. 
The lines from the second to the fifth of Fig. 6 illustrate  the case
when randomness, of increasing intensity, $\epsilon$, is superimposed to the sequence. 
We  
make $\epsilon$ vary from $10^{-6}$ to $0.1$. We see that
both CASSANDRA and CASToRe have a similar sensitivity to the 
the amount of randomness. The second line, first column, of Fig. \ref{paologiulia1} shows that even an infinitesimal
value of $\epsilon$ prevents $S(t)$ from regressing to zero. The same second line, second column, shows that the CIC value does not follow anymore a simple logarithmic increase.
As $\epsilon$ increases both CASSANDRA and CASToRe make a transition
towards the respective behavior that they are expected to show in the case of totally random signals, namely, linear increase with respect
to $log t$, the former, and with respect to $t$, the latter.

We now compare CASToRE to CASSANDRA   in a case similar to
that of Eq. \ref{modulatednoiseintensity}. 
We create  a
very long sequence with all the sites filled with $0$'s. Then with the probability $\epsilon$ we assign to any site either the symbol $0$, the symbol already there, or the symbol $1$ by tossing a coin. With the probability $1-\epsilon$ we leave the site with the original symbol $0$.
In comclusion, we realize the sequence

\begin{equation}
\xi_n=\left\{ \begin{array}{cccc} 
\mbox{random}(0,1)& \mbox{with prob.} &\epsilon(n)& \\
0 & \mbox{with prob.}&1-\epsilon(n)&. \\
\end{array}
 \right.
\end{equation}
The probability $\epsilon$ is site dependent, and we select the following modulation
\begin{equation}
\label{periodicepsilon}
\epsilon(n)=1-\cos(\Omega n).
\end{equation}

At this stage we use the version of CASToRe and CASSANDRA with two moving windows. We remind the reader that, as far as the DE method is concerned, the term CASSANDRA was coined indeed to denote the use of the DE method with the two windows. We have to establish if CASSANDRA and CASToRE perceive  correctly  the dependence of statistics on site position established  by Eq. (33).   The results
are shown in Fig. 7.  Notice that the local AIP with
a small windows of length $l=10000$ is effectively able to detect the
local amount of noise. As a result we see a local CIC$(n)$ that
evolves harmonically in time. The local CASSANDRA indicator defined by
Eq. \ref{cassandra} is not sensitive to the global amount of noise: the
$S(0)$ component of Eq.  \ref{cassandra} would be, but it is subtracted
 from the indicator $I(n)$.  $I(n)$, on the other hand, is sensitive to the
change in statistics (as we have seen in Fig. 4 ).  As stated earlier, CASSANDRA perceives
the effect of changing rules on the transition from dynamics to thermodynamics. Thus, this effect
is detectable as a function of time (or site) in the time series. In other words,  CASSANDRA
measures the rate of change of the statistics under study. This is strongly 
suggested by the fact that CASSANDRA yields a result that is the time derivative 
of that stemming from CASToRE, as clearly illustrated by Fig. 7.

\begin{figure}
\label{paologiulia2}
\includegraphics[width=7cm,angle=270]{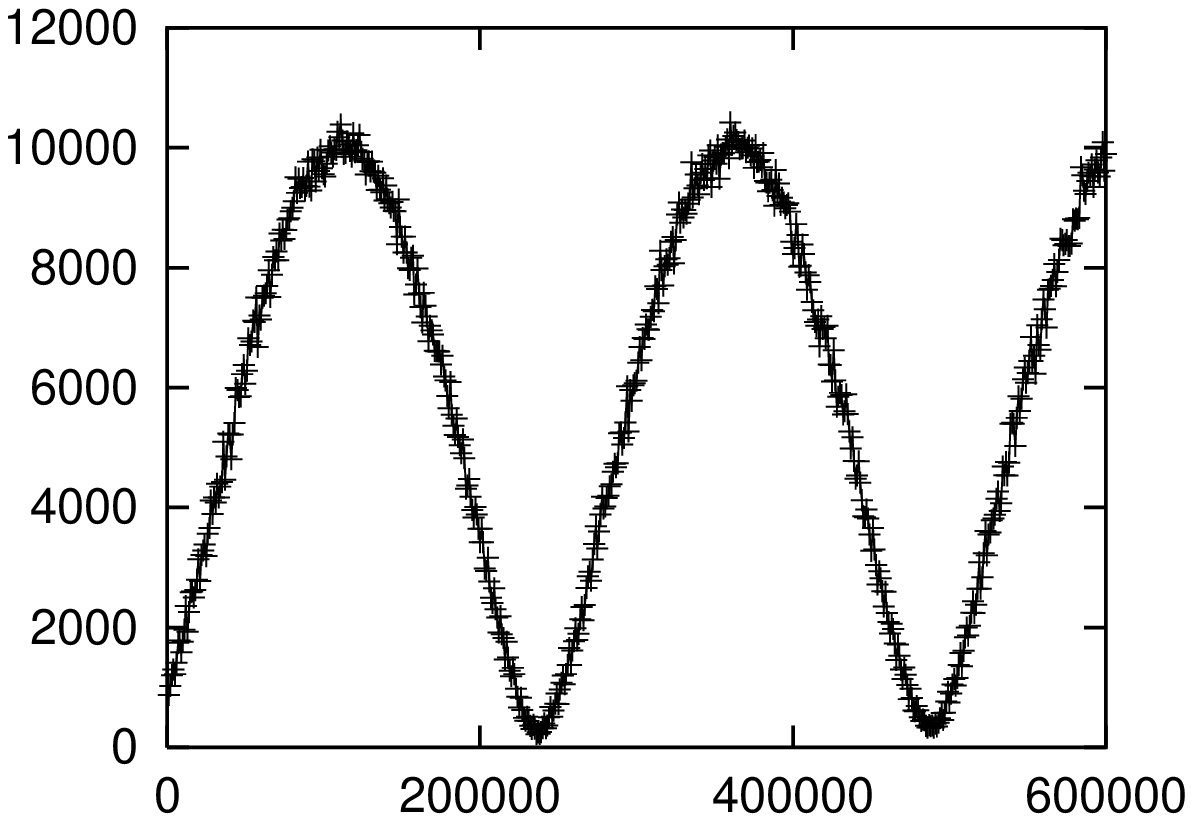}\hspace{-1cm}{\large A}
\includegraphics[width=7cm,angle=270]{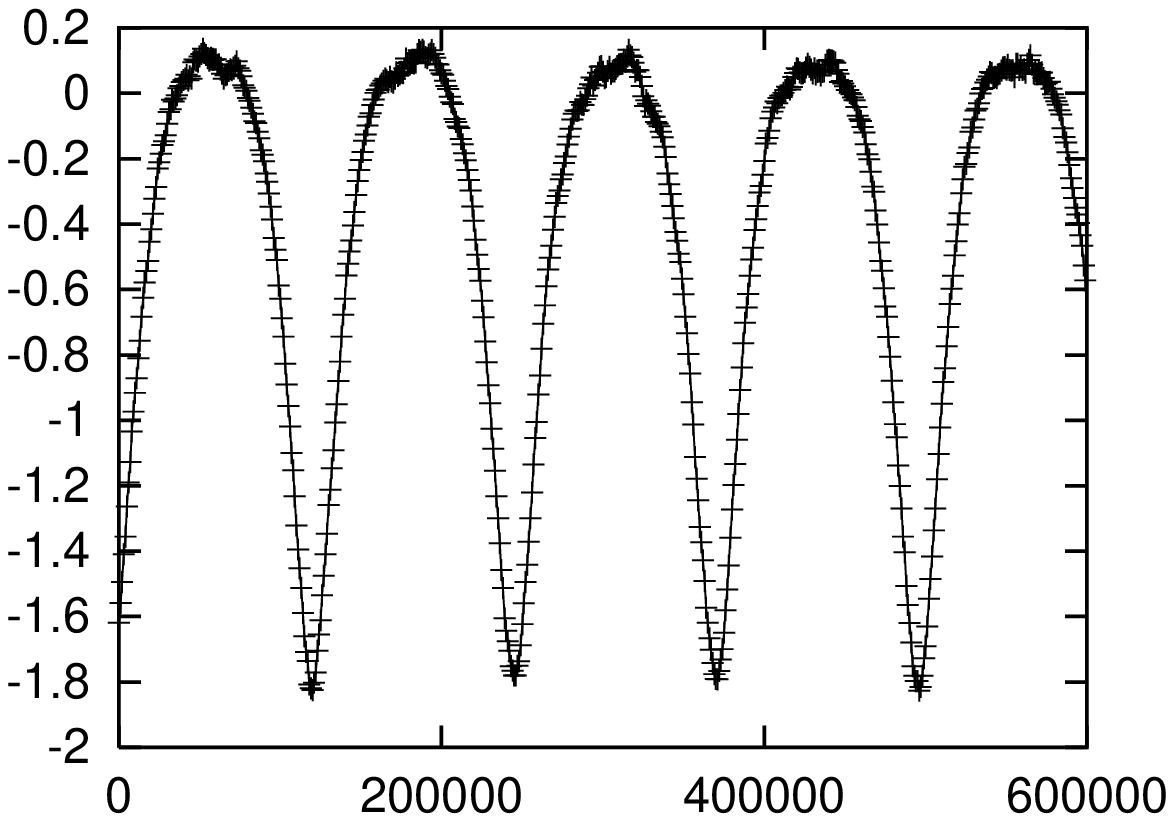}\hspace{-1cm}{\large B}
\begin{center}
\caption{Local indicators of non-random prroperties. A: the local
CIC parallels the value of $\epsilon$. B: the CASSANDRA
local indicator $I$ has its maximum value when the
derivative of $\epsilon$ with respect to time is maximum}
\end{center}
\end{figure}

\section{Events and Pseudo Events}
In Section IV we have pointed out what do we mean by event. This
definition of event, as an unpredictable random event, implies that we
regard as pseudo events the events that are, to some extent to
clarify, predictable. Let us illustrate the distinction between events
and pseudo events with the help of the dynamic model of Section IV.  Let us write it in a
more general form, as follows.
\begin{equation}
    \label{keyequation}    \dot x  =  \Phi(x) > 0,
\end{equation}
where $x$ denotes the coordinate of a particle, moving within the
interval $I \equiv [0,1]$, from the left to the right, with times of
arrival at $x = 1$ determined by Eq.(\ref{keyequation}) and by the
initial condition. When the particle reaches the right border of $I$,
it is injected back to a new initial condition selected with uniform
probability on $I$.  Consequently, the times of arrival at $x = 1$,
$t_{1},..,t_{i}..,$ represent events.  As shown in Section IV, the
choice $\Phi(x) = \kappa x^{z}$,with $z > 1$ and $k > 0$, yields for
the waiting times $\tau_{i} \equiv t_{i}- t_{i-1}$ the waiting
distribution density of Eq. (\ref{waitingtimes}).

Next, we convert the time series $\{\tau_{i}\}$ into a random walk.
We select a rule \cite{giacomo} that makes the random walker move,
always in the same direction and by a step of constant intensity, only
when an event occurs.  This means that the sequence $\{\tau_{i}\}$ is
converted into a sequence of $0$'s and $1$'s as follows.  We take the
integer part of $\tau_1$, say $n_1$, and we fill the first $n_1$ sites
of the sequence with $0$'s. We assign the symbol $1$ to the
$n_{1}+1-th$ site and we move to $\tau_2$, and so on.  The resulting
sequence is formed by attaching patches together in the same order as
the sequence of $\tau_i$. The resulting sequence has a number of
sites, say $N$, given by the sum of the integer parts of $\tau_{i}$'s.
Then, this resulting sequence is converted into many trajectories of a
given length $l$. A window of size $l$ moves along the sequence and
for any window position, the portion of the whole sequence spanned by
the window is regarded as a single trajectory of length $l$. Thus we
obtain $N + 1 -l$ trajectories that are assumed to start from the
origin, and are used to create a diffusion distribution, at time
$l$. If there is scaling, the DE method detects it by means of
Eq. (\ref{shapeofdiffusionentropy}).

The numerical evaluation of $\delta$ with the DE method
 has been recently done in Ref. \cite{nic2} and the numerical result supports the theoretical prediction according to which, in the specific case where    Eq.(\ref{waitingtimes}) applies,  
the scaling parameter $\delta$, for $2< \mu < 3$, reads

    \begin{equation} \label{nomemory} \delta = \frac{1}{\mu -1} \mbox{
	} (\delta = 0.5 \mbox{ if } \mu > 3).  \end{equation} This
	prediction is known \cite{giacomo} to be correct, only when an
	ordinary form of memory exists \cite{katia}.  A Markov master
	equation, namely a stochastic process without memory, is
	characterized by a waiting time distribution $\psi(\tau)$ with
	an exponential form.  This implies that a marked deviation
	from the exponential condition, and thus from $\delta = 0.5$,
	is a signature of the presence of memory\cite{katia}.  We
	refer to this memory as Type 1 memory.  To illustrate Type 2
	memory, we discuss a dynamic model generating both events and
	pseudo events.  For this purpose let us consider a
	two-variable model. The equation referring to the first
	variable, is given by Eq.(\ref{keyequation}), and the one
	concerning the new variable $y$, is given by

\begin{equation}
\label{keyequation2}
    \dot y = \chi(y) > 0.  \end{equation} The variables $x$ and $y$
   are the coordinates of two particles, both moving in the interval
   $I$, always from the left to the right. The initial conditions of
   the variable $y$ are always chosen randomly. The initial conditions
   of $x$, on the contrary, are not always chosen randomly, but they
   are only when the variable $y$ reaches the border at least once,
   during the sojourn of $x$ within the interval. Let us consider the
   sojourn time interval $[t_{i},t_{i+1}]$. The times $t_{i}$ signal, as in the case of the earlier 0ne-dimensonal model, the arrival of the particle of interest at $x = 1$. If in this time interval
   the variable $y$ remains within the interval, without touching the
   right border, then we set $x(t_{i+1}) = x(t_{i})$.This means that
   the next waiting time is equal to the preceding one, and
   consequently the time $t_{i+2}$, which might be predicted,
   represents a pseudo event.  A random even, or event, occurs when the randomj extraction for the initial condition
of $y$ is made. Thus, the sequence $\{t_{i}\}$ reflects
   a mixture of events and pseudo events. Let us consider the case
   where $\chi(y) = k'y^{z'}$ with $z' > 1$and $k' > 0$, so as to
   produce the power index $\mu'=z'/(z'-1)$, with $\mu' > 2$, a
   property of real events. Let us set the condition $\langle \tau
   \rangle_{x} \ll \langle \tau \rangle_{y}$.  In this case, it is
   possible to prove with intuitive arguments that the waiting time
   distribution of $x$ of Eq.(\ref{waitingtimes}) is changed into one
   much faster than the original. In fact, if we imagine a succession
   of waiting times for $y$, all equal to $<\tau>_{y}$, we see that
   the number of long waiting times for $x$ is less than the number of
   short waiting times, thereby making the perturbed waiting time
   distribution faster than the original distribution.  Let us
   consider the case where the unperturbed waiting time distribution
   $\psi(\tau)$ is characterized by $\mu > 3$ ($\mu =5$, in the case
   of Fig. 1). The perturbed waiting time distribution, to be
   identified with that experimentally observed, is even faster and
   consequently is expected to produce a diffusion process with
   $\delta = 0.5$. However, the experimental $\psi(\tau)$ is not
   simply a reflection of real events but includes pseudo events as
   well.  Fig. 8  reveals a very attractive property: the DE now
   yields $\delta = 1/(\mu'-1)$, quite different from the prescription
   of Eq. (\ref{nomemory}) that would yield $\delta = 0.5$.  The
   breakdown of Eq.(\ref{nomemory}) is a manifestation of Type 2
   memory, referred to by us as \emph{memory beyond memory}.  In fact,
   the existence of pseudo events implies correlation among different
   times of the series $\{\tau_{i}\}$, and thus a memory of earlier
   events.  The inset of Fig. 8 shows that shuffling the order
   of the corresponding patches has the effect of yielding $\delta =
   0.5$, as the experimental $\psi(\tau)$ implies.  The scaling
   detected by the DE method does not depend on the pseudo events, but
   only on the hidden events, and thus on a time distribution, which
   cannot be experimentally detected, longer than $\psi(\tau)$.

\begin{figure}[!h]
\begin{center}
\includegraphics[scale=.31]{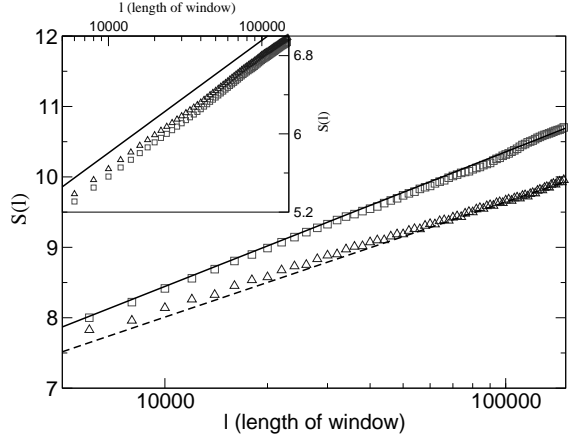}
\caption{\label{fig1} DE for two-variables model as a function of
time. The squares correspond to $k'$=0.018, $z'$=1.83 while the
diamonds to $k'$=0.011, $z'$=1.71. For both curves $k$=0.4,
$z$=1.25. Lines correspond to $\delta=0.83$ (full line) and $\delta=0.71$
(dashed line). In the inset: the same curves after shuffling, the straight
line slope is $0.5$.}
\end{center}
\end{figure}

In order to interpret this model via the CIC method, let us consider
its discrete version:
$$\left\{
\begin{array}{ll}
y_{n+1}= y_n+k'\ y_n^{z'}\Delta t &\mbox{if }0<y_n<1\\
y_{n+1}= \Xi _1 &\mbox{otherwise}\\
w_{n+1}= (1-H(y_n-1)) w_{n} + H(y_n-1)\Xi _2\\
x_{n+1}= x_n+k\ x_n^{z}\Delta t &\mbox{if }0<x_n<1\\
x_{n+1}= w_n &\mbox{otherwise}
\end{array}\right. $$ 
where $H(s)$ is a Heaviside function, $\Xi _1$ and $\Xi _2$ are two
independent random injections and the function $w_n$ is a control on
the initial position of $x_n$. Note that in the numerical calculation we set the time step
$\Delta t = 1$.

This way, the system gets an additional degree of freedom (the action
of $y_n$). Thus, we expect that the KS entropy depends only on the
$y_n$-dynamics. Hence, the resulting entropy can be compared to the
one arising from the formula (\ref{analyticalks}) (with the current
value of $z'$). The above quantity (\ref{analyticalks}) needs a factor
$2= \log _2 (2^2)$ correction, due to the increase in one degree of
freedom from the one-walker to the two-walkers model: this way we can
calculate the entropy using the theory. We have performed two numerical
simulations; the results agree with this prediction in both cases
where $z=1.25,\ k=0.4,\ z'=1.71,\ k'=0.011$ and $z=1.25,\ k=0.4,\
z'=1.83,\ k'=0.018 $. In the following table we show that the
numerical results obtained using CASToRE fit this prediction.
\begin{center}
\begin{tabular}{|c|c|c|c|c|c|}
\hline {$z$ value} & {$k$ value} & {$z'$ value} &{$k'$ value} &
{Theoretical entropy }& {CASToRe entropy}\\ \hline \hline
1.25&0.4&1.71&0.011& 0.050&0.053\\ 1.25&0.4&1.83&0.018& 0.055&0.058\\
\hline
\end{tabular}\label{table2walkers}
\end{center}
\vskip 0.5 truecm
\section{Quasiperiodic Processes}

\begin{figure}
\label{Jin}
\begin{center}
 \includegraphics[width=7cm,angle=270]{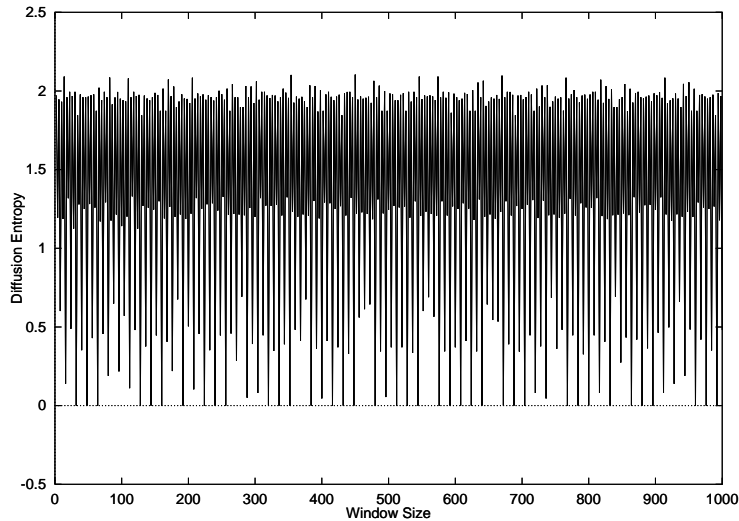}
\caption{The results of DE analysis for the logistic map at the chaos
threshold $x_{n+1}=1-\mu\ x_n^2$, where the parameter value is
$\mu=1.40115518909205$.}
\end{center}
\end{figure}
This section is devoted to illustrate a dynamic case of significant
interest, the logistic map at the chaos threshold. This case is judged
to be by many researchers as a prototype of complex systems, for which
some other researchers propose the adoption of non-additive entropic
indicator\cite{tsallisandlogisticmap}.  According to the authors of
Ref.\cite{menconi}, this case yields an increase of AIC (and CIC) proportional
to the logarithm of the window size, namely, a behavior identical to
that discussed in Section V. It is therefore of crucial importance to
check this prediction by means of CASSANDRA. We have seen that in the
periodic case CASSANDRA yields entropy saturation and recurrences,
thereby suggesting that the logarithmic increase of AIC does not have
a thermodynamic significance, and that the adoption of thermodynamic
arguments, non-extensive\cite{tsallisandlogisticmap} as well as
ordinary, does not seem to be appropriate.  It has been recently
shown \cite{simone} that the adoption of the Pesin theorem yields the
same logarithmic dependence on time of the complexity increase, in
line with the prediction of Ref.\cite{menconi}.  On the basis of all
this, we expect CASSANDRA to result in saturation and recurrences. We
see from Fig. \ref{Jin} that it is so, the only main difference with the
case discussed in Section V being that now the recurrences rather
than being regular looks erratic.

\section {Concluding remarks}

From a conceptual point of view, the most important result of this
paper is that it sheds light on the connection between scaling and
information content, and through the latter, with thermodynamics.
This is an important problem since in literature there seems to be
confusion on this intriguing issue, insofar as anomalous diffusion is
often interpreted in terms of non-ordinary thermodynamics (see, for
example \cite{tsallis}). On the contrary, we find that the
intermittent model of Section 4 yields an ordinary increase of the
information content $I$ as a function of $n$, namely, an increase
proportional to $n$ in the whole range $2 < \mu < 3$. At the diffusion
level, this corresponds to the anomalous scaling of Eq. (35). This is
the dynamical region corresponding to the emergence of L\'{e}vy
statistics \cite{allegro}. Yet, no anomaly is revealed by the
measurement of $I$ carried out by means of CASToRE. This seems to fit
the conclusion that, after all, L\'{e}vy statistics is not
incompatible with ordinary statistical mechanics even if it implies a
striking deviation from the canonical equilibrium
disrribution \cite{campisi}. It must be pointed out that in the case
when the flight perspective is adopted \cite{giacomo,debbie} the
scaling value $\delta$ keeps obeying the prescription of Eq. (35) even
if the condition $\mu < 2$ applies. This means that in that case
CASSANDRA keeps signalling the existence of scaling even if the
exponential sensitivity is replaced by stretched exponential
sensitivity.

A more significant property seems to be given by the case when the
information content increase becomes proportional to $\log n$. In this
specific case, concerning both the periodic case of Fig. 1 and the
quasi-periodic case of Fig. 8, the DE cannot exceed a maximum value
and is characterized by either periodic (Fig.1) or quasi periodic(Fig. 8)
regressions to the initially vanishing value. This means, in other
words, again in conflict with Ref. \cite{tsallis}, that in this case
we are not in the presence of a thermodynamic regime, but rather we
are forcing a periodic or quasi-periodic process to generate
diffusion. The adoption of moving windows of size $l$ to generate
distinct trajectories corresponds, in the periodic case, to producing
many trajectories with an unknown initial condition, thereby
explaining why we observe initially an entropy increase. Then, the DE
undergoes infinitely many regressions signalling to us that this form
of entropy increase depends only on the uncertainty on initial
condition, rather than on the trajectory complexity. The
quasi-periodic case (see Fig. 8) exhibits similar properties, but the
initial regime of entropy increase does not seem to exist, in this
case.

We are not in the presence of a new thermodynamic
regime, and our conclusion should be compared to  those of Refs. \cite{sokolov,klaftersok}. These authors 
prove that strange kinetics do not conflict with ordinary thermodynamics, in the sense that strange kinetics can be comptible with canonical equlilbrium. 
In our dynamic perspective, thermodynamics means Kolmogorov complexity, and it is in this sense that we agree with the authors of Refs. \cite{sokolov,klaftersok}. We think that strange kinetics do not conflict with the view established by Kolmogorov, and pursued by Pesin, on the basis of the ordinary Shannon entropy.
We do not think that anomalous diffusion requires non-ordinary thermodynamics. Rather, anomalous diffusion seems to imply a transition from
the dynamic to the thermodynamic regime that might be exceptionally
extended in time. For this reason the adoption of the two mobile
windows of Section V is expected to afford a powerful method of
analysis of time series corresponding to real complex processes. In
fact, the two techniques, both CASToRE and CASSANDRA, can be used to
explore local time conditions. If the rules driving the dynamics
process under study, through the statistical analysis of the sequences
generated by the process, change in time, there might be no time for
the scaling (thermodynamic) regime to emerge. In this condition, the
two techniques can be used to study the transition from dynamics to
thermodynamics, a transition never completed due to the fact that the
rules change before the realization of the thermodynamic regime occurs.


\end{document}